\documentstyle[12pt]{article}

\newcommand{\be}{\begin{equation}}
\newcommand{\ee}{\end{equation}}
\newcommand{\bea}{\begin{eqnarray}}
\newcommand{\eea}{\end{eqnarray}}
\newcommand{\beaa}{\begin{eqnarray*}}
\newcommand{\eeaa}{\end{eqnarray*}}

\newcommand{\BB}{{{\rm I} \kern -2pt \rlap {\rm B} \kern +8pt}}

\advance\textheight by \topskip
\input{tcilatex}
\begin{document}

\baselineskip18pt \parindent12pt \parskip10pt

\begin{titlepage}

\begin{center}
{\Large {\bf On Darboux transformation of the supersymmetric sine-Gordon
equation }}\\\vspace{1.5in} {\large
M. Siddiq\footnote{%
mohsin\_pu@yahoo.com}\footnote{%
On study leave from PRD (PINSTECH) Islamabad, Pakistan}, M. Hassan \footnote{%
mhassan@physics.pu.edu.pk} }\vspace{0.15in} and U. Saleem\footnote{%
usman\_physics@yahoo.com}\\
{\small{\it Department of Physics,\\ University of the Punjab,\\
Quaid-e-Azam Campus,\\Lahore-54590, Pakistan.}}
\end{center}

\vspace{1cm}
\begin{abstract}
Darboux transformation is constructed for superfields of the super
sine-Gordon equation
 and the superfields of the associated linear problem. The Darboux transformation is
shown
 to be related to the super B\"{a}cklund transformation and is
further used to obtain $N$ super soliton solutions.
\end{abstract}
\vspace{1cm} \vspace{1cm} PACS: 02.30.Ik
\end{titlepage}

There has been an increasing interest in the study of supersymmetric
integrable systems for the last few decades \cite{Kuper}-\cite{usman}. Among
the many techniques used to study integrability and to obtain the
multisoliton solutions for a given integrable model, Darboux transformation%
\footnote{%
Originally the Darboux transformation was first introduced by Darboux back
in $1882$, in the study of pseudo-spherical surfaces and later used to
generate solutions of the Sturm-Liouville differential equation. The Darboux
transformation has now widely been used to generate multisoliton solutions
of integrable as well as super integrable evolution equations.} has been
widely used and it has established itself as an economic, convenient and
efficient way of generating solutions \cite{matveev}-\cite{darb}. The
Darboux transformation has been employed on some supersymmetric integrable
models in recent years \cite{liu}-\cite{liu4}. In these investigations
multisoliton solutions have been constructed and the ideas are generalized
to incorporate the Crum transformation, the Wronskian superdeterminant and
Pfaffian type solutions. The super soliton solutions of the super KdV
equation and super sine-Gordon equation have been investigated and it has
been shown that the solitons of the KdV and sine-Gordon solitons appear as
the body of the super solitons \cite{liu1}-\cite{liu}.

The purpose of this work is to provide a thorough investigation of Darboux
transformation for super sine-Gordon equation in a systematic way and to
obtain the explicit super multisoliton solutions by a Crum type
transformation. Following \cite{Siddiq}, we write a linear problem in
superspace whose compatibility condition is the super sine-Gordon equation.
The linear problem then leads to a Lax formalism in superspace. We
explicitly write the Darboux transformation for the fermionic and bosonic
superfields of the linear system and for the scalar superfield of the super
sine-Gordon equation. The approach adopted here is different from that
adopted in \cite{liu4}. In ref \cite{liu4} the authors have not explicitly
constructed the Darboux transformation and $N$-soliton solutions for the
super sine-Gordon equation. We have extended the results of \cite{liu4} to
give $N$ super soliton solutions of the super sine-Gordon equation in terms
of known solutions of the linear problem and express it as a series of
products of super determinants and fermionic superfields. The Darboux
transformation is then shown to be related to the super B\"{a}cklund
transformation of the super sine-Gordon equation \cite{Ku}.

We follow the general procedure of writing manifestly supersymmetric
sine-Gordon equation. The equation is defined in two dimensional
super-Minkowski space with bosonic light-cone coordinates $x^{\pm }\footnote{%
Our space-time conventions are such that the orthonormal and light-cone
coordinates are related by $x^{\pm }=\frac{1}{2}(t\pm x)$ and $\partial
_{\pm }=\frac{1}{2}(\partial _{t}\pm \partial _{x})$.}$ and fermionic
coordinates $\theta ^{\pm }$, which are Majorana spinors. The superspace
lagrangian density of $N=1$ super sine-Gordon theory is given by 
\begin{equation}
{\cal L}(\Phi )=\frac{i}{2}D_{+}\Phi D_{-}\Phi +\cos \Phi ,  \label{lag1}
\end{equation}%
where $\Phi $ is a real scalar superfield and $D_{\pm }$ are covariant
superspace derivatives defined as 
\[
D_{\pm }=\frac{\partial }{\partial \theta ^{\pm }}-i\theta ^{\pm }\partial
_{\pm },\,\,\,\,\,\,\,D_{\pm }^{2}=-i\partial _{\pm
},\,\,\,\,\,\,\,\,\,\{D_{+},D_{-}\}=0, 
\]%
where $\{\,,\,\}$ is an anti-commutator. The superfield evolution equation
follows from the lagrangian and is given by 
\begin{equation}
D_{+}D_{-}\Phi =i\sin \Phi .  \label{sin}
\end{equation}%
The equation (\ref{sin}) is invariant under $N=1$ supersymmetry
transformations.

Let us first recall that the super sine-Gordon equation appears as the
compatibility condition of the following linear system of equations \cite%
{Siddiq}

\begin{equation}
D_{\pm }\Psi ={\cal A}_{\pm }{\cal \,}\Psi ,  \label{lin1}
\end{equation}%
where the superfield $\Psi $ is expressed in terms of bosonic and fermionic
superfield components as 
\[
\Psi =\left( 
\begin{array}{l}
\psi \\ 
\phi \\ 
\chi%
\end{array}%
\right) ,\qquad 
\]%
where $\psi $, $\phi $ are bosonic superfields and $\chi $ is a fermionic
superfield and ${\cal A}_{\pm }$ are $3\times 3$ matrices given by 
\begin{eqnarray*}
{\cal A}_{+} &=&\frac{1}{2\sqrt{\lambda }}\left( 
\begin{array}{lll}
\,\,\,\,\,\,\,\,\,\,0 & \,0 & \,\,\,\,\,\,\,\,\,\,\,\,\,\,\,\,\,i\,e^{i\Phi }
\\ 
\,\,\,\,\,\,\,\,\,\,0 & \,0\, & \,\,\,\,\,\,\,\,-\,i\,\,e^{-i\Phi } \\ 
\,\,-\,e^{-i\Phi } & e^{i\Phi } & \,\,\,\,\,\,\,\,\,\,\,\,\,\,\,\,\,\,\,\,\,0%
\end{array}%
\right) , \\
{\cal A}_{-} &=&\sqrt{\lambda }\left( 
\begin{array}{lll}
\,\,\,\,\,\,\,\,\,\,\frac{i\,D_{-}\Phi }{\sqrt{\lambda }} & 
\,\,\,\,\,\,\,\,\,\,\ \ 0 & \,\,~\ -i \\ 
\,\,\,\,\,\,\,\,\,\ \ \ \,0 & \,-\frac{i\,D_{-}\Phi }{\sqrt{\lambda }}\, & \
\ \ \ \ i \\ 
\ \ \ \ -1 & \ \ \ \ \ 1 & \,\,\,\,\,\,\,\,\,0%
\end{array}%
\right) .
\end{eqnarray*}%
The zero curvature condition in superspace 
\begin{equation}
D_{+}{\cal A}_{-}+D_{-}{\cal A}_{+}-\{{\cal A}_{+},{\cal A}_{-}\}=0,
\label{zero1}
\end{equation}%
gives the equation (\ref{sin}). The zero curvature condition (\ref{zero1})
is essentially the compatibility condition of the linear system (\ref{lin1}%
). As in the bosonic case, the zero curvature condition of the super
sine-Gordon equation gives a Lax representation; one can directly find the
bosonic Lax operator for the supersymmetric case such that the given
operators obey the Lax equation. The Lax equation gives the evolution of the
spectral problem and solves the super sine-Gordon equation in the spirit of
the inverse scattering method \footnote{%
We can also have a gauge equivalent description of the linearization and the
Lax operator solves the isospectral problem for the gauge transformed
eigenfunctions.} \cite{Siddiq}.

The Darboux transformation is used to generate multisoliton solutions of
integrable as well as super integrable evolution equations. Given a
superfield equation (\ref{sin}) and its associated linear system (\ref{lin1}%
), one can construct a Darboux transformation on the superfields $\psi $, $%
\phi $ and $\chi $ of the linear system (\ref{lin1}) and on the sine-Gordon
superfield $\Phi $ such that the new transformed fields obey the same
differential equations for the value of spectral parameter $\lambda =\lambda
_{0}$. The one-fold Darboux transformation for the system (\ref{lin1}) is
given by 
\begin{eqnarray}
\psi (\lambda _{0}) &\rightarrow &\psi \lbrack 1](\lambda _{0};\lambda
_{1})=\lambda _{0}\phi (\lambda _{0})-\lambda _{1}\frac{\phi _{1}(\lambda
_{1})}{\psi _{1}(\lambda _{1})}\psi (\lambda _{0})-i\sqrt{\lambda
_{0}\lambda _{1}}\frac{\chi _{1}(\lambda _{1})}{\psi _{1}(\lambda _{1})}\chi
(\lambda _{0}),  \label{dt0} \\
\phi (\lambda _{0}) &\rightarrow &\phi \lbrack 1](\lambda _{0};\lambda
_{1})=\lambda _{0}\psi (\lambda _{0})-\lambda _{1}\frac{\psi _{1}(\lambda
_{1})}{\phi _{1}(\lambda _{1})}\phi (\lambda _{0})-i\sqrt{\lambda
_{0}\lambda _{1}}\frac{\chi _{1}(\lambda _{1})}{\phi _{1}(\lambda _{1})}\chi
(\lambda _{0}),  \nonumber \\
\chi (\lambda _{0}) &\rightarrow &\chi \lbrack 1](\lambda _{0};\lambda
_{1})=-(\lambda _{0}+\lambda _{1})\chi (\lambda _{0})+\sqrt{\lambda
_{0}\lambda _{1}}\frac{\chi _{1}(\lambda _{1})}{\psi _{1}(\lambda _{1})}\psi
(\lambda _{0})+\sqrt{\lambda _{0}\lambda _{1}}\frac{\chi _{1}(\lambda _{1})}{%
\phi _{1}(\lambda _{1})}\phi (\lambda _{0}),  \nonumber
\end{eqnarray}%
where $\psi (\lambda _{0})$, $\phi (\lambda _{0})$ and $\chi (\lambda _{0})$
are the solutions of the system (\ref{lin1}) with $\lambda =\lambda _{0}$
and $\psi _{1}(\lambda _{1})$, $\phi _{1}(\lambda _{1})$ and $\chi
_{1}(\lambda _{1})$ are solution of the system (\ref{lin1}) with $\lambda
=\lambda _{1}$. The functions $\ \psi \lbrack 1](\lambda _{0};\lambda _{1})$%
, $\phi \lbrack 1](\lambda _{0};\lambda _{1})$ and $\chi \lbrack 1](\lambda
_{0};\lambda _{1})$ are solutions of the system (\ref{lin1}) with $\ \Phi $
transforming as 
\begin{equation}
\Phi (\lambda _{0})\rightarrow \Phi \lbrack 1](\lambda _{0};\lambda
_{1})=\Phi (\lambda _{0})+i\ln \frac{\psi _{1}(\lambda _{1})}{\phi
_{1}(\lambda _{1})}.  \label{dt}
\end{equation}%
where $\Phi \lbrack 1](\lambda _{0};\lambda _{1})$ is a new solution of
equation ( \ref{sin}) provided that the superfields $\psi $, $\phi $ and $%
\chi $ transform according to the transformations given in (\ref{dt0}).

Similarly one can have the following transformations 
\begin{eqnarray}
e^{i\Phi (\lambda _{0})} &\rightarrow &e^{i\Phi \lbrack 1](\lambda
_{0};\lambda _{1})}=e^{i\Phi (\lambda _{0})}\frac{\phi _{1}(\lambda _{1})}{%
\psi _{1}(\lambda _{1})},  \label{dt1} \\
D_{-}\Phi (\lambda _{0}) &\rightarrow &D_{-}\Phi \lbrack 1](\lambda
_{0};\lambda _{1})=-D_{-}\Phi (\lambda _{0})+\sqrt{\lambda _{1}}\left[ \frac{%
\chi _{1}(\lambda _{1})}{\psi _{1}(\lambda _{1})}+\frac{\chi _{1}(\lambda
_{1})}{\phi _{1}(\lambda _{1})}\right] ,  \label{dt2} \\
D_{+}\Phi (\lambda _{0}) &\rightarrow &D_{+}\Phi \lbrack 1](\lambda
_{0};\lambda _{1})=D_{+}\Phi (\lambda _{0})-  \nonumber \\
&&\frac{1}{2\sqrt{\lambda _{1}}}\left[ e^{i\Phi (\lambda _{0})}\frac{\chi
_{1}(\lambda _{1})}{\psi _{1}(\lambda _{1})}+e^{-i\Phi (\lambda _{0})}\frac{%
\chi _{1}(\lambda _{1})}{\phi _{1}(\lambda _{1})}\right] .  \label{dt3}
\end{eqnarray}%
In these transformations, $\Phi \lbrack 1]$ is another solution of the super
sine-Gordon equation (\ref{sin}) with $\lambda =\lambda _{0}$ generated
through the Darboux transformation. In fact the solution $\Phi \lbrack 1]$
is expressed in terms of solution $\Phi (\lambda _{0})$ and particular
solutions $\psi _{1}(\lambda _{1})$ and $\phi _{1}(\lambda _{1})$ of the
linear system (\ref{lin1}) with $\lambda =\lambda _{1}$.

To establish a connection between Darboux transformation and B\"{a}cklund
transformation of the super sine-Gordon equation (\ref{sin}), we write $%
\Gamma =\frac{\psi _{1}}{\phi _{1}}$ and $k=\frac{\chi _{1}}{\psi _{1}}$ and
express equation (\ref{dt2}) in terms of $\Gamma $ and $k$ as 
\[
D_{-}\Phi \lbrack 1]=-D_{-}\Phi +\sqrt{\lambda _{1}}\left[ k+k\Gamma \right]
. 
\]%
Now eliminating $\Gamma $ and $k$ by a transformation $\Gamma =\exp i(\Phi
-\Phi \lbrack 1])$ and $f=k\sqrt{\Gamma }$, we have

\begin{equation}
D_{-}(\Phi +\Phi \lbrack 1])=2\sqrt{\lambda _{1}}\,\,f\,\,\cos \left( \frac{%
\Phi -\Phi \lbrack 1]}{2}\right) ,  \label{bt1}
\end{equation}%
where $f$ is another fermionic superfield. In the same manner we can find
the other part of the B\"{a}cklund transformation, from equation (\ref{dt3}) 
\begin{equation}
D_{+}(\Phi -\Phi \lbrack 1])=\frac{1}{\sqrt{\lambda _{1}}}\,\,f\,\,\cos
\left( \frac{\Phi +\Phi \lbrack 1]}{2}\right) .  \label{bt2}
\end{equation}%
The superfield $f$ is subjected to the following conditions 
\begin{eqnarray}
D_{+}f &=&\frac{i}{\sqrt{\lambda _{1}}}\sin \left( \frac{\Phi +\Phi \lbrack
1]}{2}\right) ,  \label{bt3} \\
D_{-}f &=&-2i\sqrt{\lambda _{1}}\sin \left( \frac{\Phi -\Phi \lbrack 1]}{2}%
\right) .  \label{bt4}
\end{eqnarray}%
Equations (\ref{bt1})-(\ref{bt4}) are exactly the B\"{a}cklund
transformation of the super sine-Gordon equation already obtained in \cite%
{Ku}. The super soliton solution can be obtained by taking $\Phi =0$ in (\ref%
{dt}) to give a solution of the linear problem (\ref{lin1}) as 
\begin{eqnarray*}
\psi (x^{\pm },\theta ^{\pm }) &=&A_{0}^{+}(\theta ^{+},\theta ^{-})\exp
(\eta )+A_{0}^{-}(\theta ^{+},\theta ^{-})\exp (-\eta ), \\
\phi (x^{\pm },\theta ^{\pm }) &=&A_{0}^{+}(\theta ^{+},\theta ^{-})\exp
(\eta )-A_{0}^{-}(\theta ^{+},\theta ^{-})\exp (-\eta ), \\
\chi (x^{\pm },\theta ^{\pm }) &=&A_{1}(\theta ^{+},\theta ^{-})\exp (\eta ),
\end{eqnarray*}%
where $A_{0}^{\pm }(\theta ^{+},\theta ^{-})$ are even supernumbers and $%
A_{1}(\theta ^{+},\theta ^{-})$ is an odd supernumber and $\eta =\frac{1}{%
4\lambda }x^{+}+\lambda \,x^{-}$. By substituting this, one can obtain a
single super soliton solution as 
\[
\Phi \lbrack 1]=i\,\ln \left[ \frac{1+{\cal A}_{0}(\theta ^{+},\theta
^{-})\exp (-2\eta )}{1-{\cal A}_{0}(\theta ^{+},\theta ^{-})\exp (-2\eta )}%
\right] , 
\]%
where ${\cal A}_{0}(\theta ^{+},\theta ^{-})$ is some even supernumber. The
two super soliton solution can be obtained from the two-fold Darboux
transformation as 
\begin{eqnarray*}
\Phi \lbrack 2] &=&\Phi \lbrack 1]+i\ln \frac{\psi _{1}[1]}{\phi _{1}[1]} \\
&=&i\,\ln \left[ \frac{\Delta _{12}^{1}[2]+i\sqrt{\lambda _{1}\lambda _{2}}%
X_{12}}{\Delta _{12}^{2}[2]+i\sqrt{\lambda _{1}\lambda _{2}}X_{12}}\right] ,
\end{eqnarray*}%
where 
\begin{eqnarray*}
\Delta _{12}^{1}[2] &=&\det \left( 
\begin{array}{ll}
\lambda _{1}\phi _{1} & \lambda _{2}\phi _{2} \\ 
\psi _{1} & \psi _{2}%
\end{array}%
\right) , \\
\Delta _{12}^{2}[2] &=&\det \left( 
\begin{array}{ll}
\lambda _{1}\psi _{1} & \lambda _{2}\psi _{2} \\ 
\phi _{1} & \phi _{2}%
\end{array}%
\right) , \\
X_{12} &=&\chi _{1}\chi _{2}.
\end{eqnarray*}

Similarly the three super soliton solution as obtained by using the three
fold Darboux transformation, is given by 
\begin{eqnarray*}
\Phi \lbrack 3] &=&\Phi \lbrack 2]+i\ln \frac{\psi _{1}[2]}{\phi _{1}[2]} \\
&=&i\,\ln \left[ \frac{%
\begin{array}{c}
\Delta _{123}^{1}[3]+i\sqrt{\lambda _{2}\lambda _{3}}(\lambda _{1}+\lambda
_{2})(\lambda _{1}+\lambda _{3})\Delta _{1}^{1}[1]\,X_{23}+ \\ 
i\sqrt{\lambda _{1}\lambda _{3}}(\lambda _{1}+\lambda _{2})(\lambda
_{2}+\lambda _{3})\Delta _{2}^{1}[1]\,X_{13}+ \\ 
i\sqrt{\lambda _{1}\lambda _{2}}(\lambda _{1}+\lambda _{3})(\lambda
_{2}+\lambda _{3})\Delta _{3}^{1}[1]\,X_{12}%
\end{array}%
}{%
\begin{array}{c}
\Delta _{123}^{2}[3]+i\sqrt{\lambda _{2}\lambda _{3}}(\lambda _{1}+\lambda
_{2})(\lambda _{1}+\lambda _{3})\Delta _{1}^{2}[1]\,X_{23}+ \\ 
i\sqrt{\lambda _{1}\lambda _{3}}(\lambda _{1}+\lambda _{2})(\lambda
_{2}+\lambda _{3})\Delta _{2}^{2}[1]\,X_{13}+ \\ 
i\sqrt{\lambda _{1}\lambda _{2}}(\lambda _{1}+\lambda _{3})(\lambda
_{2}+\lambda _{3})\Delta _{3}^{2}[1]\,X_{12}%
\end{array}%
}\right] ,
\end{eqnarray*}%
where the determinants in the above expressions are given by 
\begin{eqnarray*}
\Delta _{123}^{1}[3] &=&\det \left( 
\begin{array}{lll}
\lambda _{1}^{2}\psi _{1} & \lambda _{2}^{2}\psi _{2} & \lambda _{3}^{2}\psi
_{3} \\ 
\lambda _{1}\phi _{1} & \lambda _{2}\phi _{2} & \lambda _{3}\phi _{3} \\ 
\psi _{1} & \psi _{2} & \psi _{3}%
\end{array}%
\right) , \\
\Delta _{123}^{2}[3] &=&\det \left( 
\begin{array}{lll}
\lambda _{1}^{2}\phi _{1} & \lambda _{2}^{2}\phi _{2} & \lambda _{3}^{2}\phi
_{3} \\ 
\lambda _{1}\psi _{1} & \lambda _{2}\psi _{2} & \lambda _{3}\psi _{3} \\ 
\phi _{1} & \phi _{2} & \phi _{3}%
\end{array}%
\right) , \\
\Delta _{l}^{1}[1] &=&\psi _{l}, \\
\Delta _{l}^{2}[1] &=&\phi _{l}, \\
X_{ij} &=&\chi _{i}\chi _{j}.
\end{eqnarray*}%
The super $N$-soliton solution is obtained by iteration to $N$-fold Darboux
transformation 
\begin{equation}
\Phi \lbrack N]=i\,\ln \left[ \frac{%
\begin{array}{c}
\,\dsum\limits\begin{Sb} \text{all~possible~} \\ \text{pairings~}M \end{Sb} %
\dsum\limits_{k_{1}<k_{2}<\cdots <k_{2M}}(i)^{M}\sqrt{\lambda
_{k_{1}}\lambda _{k_{2}}\cdots \lambda _{k_{2M}}}P_{12\cdots \hat{k}_{1}\hat{%
k}_{2}\cdots \hat{k}_{2M}\cdots N}(N-2M)~ \\ 
\,\left[ \Delta _{12\cdots \hat{k}_{1}\hat{k}_{2}\cdots \hat{k}_{2M}\cdots
N}^{1}[N-2M]+\right.  \\ 
\left. \frac{(i)^{M}}{M!}\sqrt{\lambda _{1}\cdots \hat{\lambda}_{k_{1}}\hat{%
\lambda}_{k_{2}}\cdots \hat{\lambda}_{k_{2M}}\cdots \lambda _{N}}%
\,X_{12\cdots \hat{k}_{1}\hat{k}_{2}\cdots \hat{k}_{2M}\cdots N}[N-2M]\right]
\,\,\,\,X_{k_{1}k_{2}\cdots k_{2M}}\,[2M]\,\,\,\,\,\,%
\end{array}%
}{%
\begin{array}{c}
\,\dsum\limits\begin{Sb} \text{all~possible~} \\ \text{pairings~}M \end{Sb} %
\dsum\limits_{k_{1}<k_{2}<\cdots <k_{2M}}(i)^{M}\sqrt{\lambda
_{k_{1}}\lambda _{k_{2}}\cdots \lambda _{k_{2M}}}P_{12\cdots \hat{k}_{1}\hat{%
k}_{2}\cdots \hat{k}_{2M}\cdots N}(N-2M)~ \\ 
\,\left[ \Delta _{12\cdots \hat{k}_{1}\hat{k}_{2}\cdots \hat{k}_{2M}\cdots
N}^{2}[N-2M]+\right.  \\ 
\left. \frac{(i)^{M}}{M!}\sqrt{\lambda _{1}\cdots \hat{\lambda}_{k_{1}}\hat{%
\lambda}_{k_{2}}\cdots \hat{\lambda}_{k_{2M}}\cdots \lambda _{N}}%
\,X_{12\cdots \hat{k}_{1}\hat{k}_{2}\cdots \hat{k}_{2M}\cdots N}[N-2M]\right]
\,\,\,\,X_{k_{1}k_{2}\cdots k_{2M}}\,[2M]%
\end{array}%
}\right] ,  \label{formula1}
\end{equation}%
where $0\leq M\leq \frac{N}{2}.$ In the formula (\ref{formula1}), the caret
above a term or an index means that it is to be omitted from the product and
the sum runs over all permutations. If $k_{1}k_{2}\cdots k_{2M}$ are omitted
indices and $12\cdots N-2M$ are unomitted indices then the polynomial $%
P_{12\cdots \hat{k}_{1}\hat{k}_{2}\cdots \hat{k}_{2M}\cdots N}(N-2M)~$is
given by 
\begin{eqnarray*}
P_{12\cdots \hat{k}_{1}\hat{k}_{2}\cdots \hat{k}_{2M}\cdots N}(N-2M)~
&=&(\lambda _{1}+\lambda _{k_{1}})(\lambda _{1}+\lambda _{k_{2}})\cdots
(\lambda _{1}+\lambda _{k_{2M}}) \\
&&(\lambda _{2}+\lambda _{k_{1}})(\lambda _{2}+\lambda _{k_{2}})\cdots
(\lambda _{2}+\lambda _{k_{2M}})\cdots  \\
&&\cdots (\lambda _{N}+\lambda _{k_{1}})(\lambda _{N}+\lambda
_{k_{2}})\cdots (\lambda _{N}+\lambda _{k_{2M}}) \\
\text{with }P_{12\cdots \cdots N}(N) &=&P_{\hat{1}~\hat{2}~\cdots \cdots 
\hat{N}}(0)=1.
\end{eqnarray*}%
The other components in the formula (\ref{formula1}) are defined as

\begin{eqnarray*}
\Delta _{12\cdots N}^{1}[N] &=&\det \left( 
\begin{array}{lllll}
\lambda _{1}^{N-1}\psi _{1} & \lambda _{2}^{N-1}\psi _{2} & \lambda
_{3}^{N-1}\psi _{3} & \cdots  & \lambda _{N}^{N-1}\psi _{N} \\ 
\lambda _{1}^{N-2}\phi _{1} & \lambda _{2}^{N-2}\phi _{2} & \lambda
_{3}^{N-2}\phi _{3} & \cdots  & \lambda _{N}^{N-2}\phi _{N} \\ 
\vdots  & \vdots  & \ddots  & \vdots  & \vdots  \\ 
\lambda _{1}\phi _{1} & \lambda _{2}\phi _{2} & \lambda _{3}\phi _{3} & 
\cdots  & \lambda _{N}\phi _{N} \\ 
\psi _{1} & \psi _{2} & \psi _{3} & \cdots  & \psi _{N}%
\end{array}%
\right) , \\
\Delta _{12\cdots N}^{2}[N] &=&\det \left( 
\begin{array}{lllll}
\lambda _{1}^{N-1}\phi _{1} & \lambda _{2}^{N-1}\phi _{2} & \lambda
_{3}^{N-1}\phi _{3} & \cdots  & \lambda _{N}^{N-1}\phi _{N} \\ 
\lambda _{1}^{N-2}\psi _{1} & \lambda _{2}^{N-2}\psi _{2} & \lambda
_{3}^{N-2}\psi _{3} & \cdots  & \lambda _{N}^{N-2}\psi _{N} \\ 
\vdots  & \vdots  & \ddots  & \vdots  & \vdots  \\ 
\lambda _{1}\psi _{1} & \lambda _{2}\psi _{2} & \lambda _{3}\psi _{3} & 
\cdots  & \lambda _{N}\psi _{N} \\ 
\phi _{1} & \phi _{2} & \phi _{3} & \cdots  & \phi _{N}%
\end{array}%
\right) , \\
X_{12\cdots \hat{k}_{1}\hat{k}_{2}\cdots \hat{k}_{2M}\cdots N}[N-2M] &=&\chi
_{1}\chi _{2}\cdots \hat{\chi}_{k_{1}}\hat{\chi}_{k_{2}}\cdots \hat{\chi}%
_{k_{2M}}\cdots \chi _{N}\text{,}~\text{with }[N-2M]\text{ even,} \\
X_{k_{1}k_{2}\cdots k_{2M}}\,[2M]\,\,\, &=&\chi _{k_{1}}\chi _{k_{2}}\cdots
\chi _{k_{2M}},
\end{eqnarray*}%
in the formula (\ref{formula1}), the term $X_{12\cdots \hat{k}_{1}\hat{k}%
_{2}\cdots \hat{k}_{2M}\cdots N}[N-2M]$ will only appear when $N$ is even.

In summary, we have investigated Darboux transformation for the super
sine-Gordon equation and obtained explicit transformations on the
sine-Gordon superfield as well as on the superfields of the linear system
associated with the equation. We have also established a connection between
the super Darboux transformation and the super B\"{a}cklund transformation
of the equation. At the end of the paper, we have used the Darboux
transformation to obtain $N$ super soliton solutions of the equation and the
solution appears in the form of product of Wronskian determinants and even
number of fermionic superfields.

{\Large Acknowledgement}

The authors acknowledge the enabling role of the Higher Education Commission
Islamabad, Pakistan and appreciate its financial support through
\textquotedblleft Merit Scholarship Scheme for Ph.D. studies in Science \&
Technology (200 Scholarships)\textquotedblright .

\end{document}